\documentclass[aps,prx,reprint,superscriptaddress,amsmath,floatfix,longbibliography]{revtex4-2}

\usepackage[utf8]{inputenc}
\usepackage[T1]{fontenc}
\usepackage{newtxtext,newtxmath}
\usepackage{graphicx}
\usepackage{booktabs}
\usepackage{siunitx}
\usepackage{chemformula}
\usepackage{multirow}
\usepackage{epstopdf}
\usepackage{tikz}
\usetikzlibrary{decorations.markings}
\usepackage{xcolor}

\usepackage[section]{placeins}

\usepackage{xurl}
\usepackage{hyperref}
\usepackage{cleveref}
\hypersetup{colorlinks=true,linkcolor=blue!45!black,citecolor=blue!45!black,urlcolor=blue!45!black}
\crefname{figure}{Fig.}{Figs.}
\Crefname{figure}{Fig.}{Figs.}

\newcommand{\vS}{\mathbf{S}}

\begin{document}

\title{Ab Initio Approach for Quantum Magnets with Coupled Cluster and Transcorrelation}

\author{Kristoffer Simula}
\email[Contact author: ]{k.simula@fkf.mpg.de}
\affiliation{Max Planck Institute for Solid State Research, Heisenbergstra{\ss}e 1, 70569 Stuttgart, Germany}
\author{Nikolay A. Bogdanov}
\affiliation{Max Planck Institute for Solid State Research, Heisenbergstra{\ss}e 1, 70569 Stuttgart, Germany}
\author{Olle Gunnarsson}
\affiliation{Max Planck Institute for Solid State Research, Heisenbergstra{\ss}e 1, 70569 Stuttgart, Germany}
\author{Daniel Kats}
\affiliation{Max Planck Institute for Solid State Research, Heisenbergstra{\ss}e 1, 70569 Stuttgart, Germany}
\author{Ali Alavi}
\affiliation{Max Planck Institute for Solid State Research, Heisenbergstra{\ss}e 1, 70569 Stuttgart, Germany}
\affiliation{Yusuf Hamied Department of Chemistry, University of Cambridge, Lensfield Road, Cambridge CB2 1EW, United Kingdom}

\date{\today}

\begin{abstract}
The nearest-neighbour exchange coupling $J_1$ is the simplest ingredient of model Hamiltonians used to understand magnetism in real materials. Nevertheless, its \textit{ab initio} determination requires a high-level correlation treatment and a large basis for wave function expansion. Here we compute $J_1$ for five oxygen-bridged spin-$\frac{1}{2}$ quantum magnets --- Sr$_2$CuO$_3$, La$_2$CuO$_4$, NdNiO$_2$, ZnCu$_3$(OH)$_6$Cl$_2$, and Li$_2$CuO$_2$ --- spanning bridging angles from 180$^\circ$ to 94$^\circ$ and the crossover from antiferromagnetic to ferromagnetic coupling, using embedded-cluster methodology. To describe the open-shell singlets we use 2-determinant coupled cluster with single and double excitations, which recovers the orbital-breathing enhancement of superexchange from a minimal reference. Basis-set errors are found to be large: extrapolation to the complete basis-set (CBS) limit shifts $J_1$ by up to 50~meV from bases of double-$\zeta$ quality. Transcorrelated Hamiltonians reproduce the CBS estimates with modest basis set sizes, yielding an excellent accuracy against experiment at highly affordable cost. These results are the first application of transcorrelated methods in strongly correlated solid-state systems, and establish transcorrelation as a powerful theory in computational research of materials.
\end{abstract}

\maketitle

\section{Introduction}

Materials that have low-dimensional, strongly correlated magnetic sublattices are a rich source of exotic physics and widely studied for their use as a platform for new theory and technological applications. Especially materials with oxygen-mediated exchange between magnetic sites, i.e., with superexchange, have been of much interest. Among other phenomena, they show Mott and charge-transfer insulating states, can achieve high-temperature superconducting phases via doping, and frustrated magnetic sublattices can host complex spin liquid state \cite{lee2006, shores2005, Khomskii_2014}. A complete description of correlated oxide materials remains out of reach, and even making the connection between experimental measurements and theoretical results is not a trivial task. 

In cuprate Mott insulators with 3d$^9$ configuration, transition-metal ions carry a single hole in the $x^2\!-\!y^2$ orbital, and the half-filled lattice can remain insulating due to the large on-site Coulomb repulsion $U$. Virtual electron hopping with amplitude $t$ between neighbouring sites with antiparallel spins --- forbidden for parallel spins by the Pauli principle --- yields an energy gain of order $4t^2/U$, defining an antiferromagnetic (AFM) exchange coupling $J_1>0$. Well-known Goodenough-Kanamori rules~\cite{goodenough1955, kanamori1959} state that this superexchange is strongest for a straight 180$^\circ$ Cu--O--Cu (or Ni--O--Ni) bridge; as the bridge bends towards 90$^\circ$, symmetry suppresses the ligand-mediated hopping, and a ferromagnetic (FM) channel driven by Hund's coupling on the bridging ligand emerges in its place, leading to $J_1<0$~\cite{mizuno1998}. In frustrated lattices no long-range order sets in, but $J_1$ remains the key parameter of the underlying Heisenberg model \cite{shores2005,rigol2007}. The infinite-layer nickelate NdNiO$_2$ has with Ni$^+$ ions a d$^9$ occupancy in 2-dimensional magnetic sublattices, and superconducts upon doping; its dispersion relation resembles that of a spin-$1/2$ antiferromagnet and has been considered to be weakly metallic~\cite{lu2021}.

The understanding of exotic phases of matter is in large part built on model Hamiltonians, such as the Hubbard Hamiltonian. However, the choice of model and the determination of the corresponding parameters is often a nontrivial task. In particular, some model parameters may not be related to experimental results in a simple way. Their values depend on the amount of physics explicitly built into the model, and the physics not included explicitly should ideally be included implicitly by using renormalized parameters. \textit{Ab initio} calculations can give guidance in the choice of an appropriate model and can help establishing relations between model parameters and experimental results. Systematically improvable wave-function-based quantum chemistry methods are very useful in this context. They are applicable to strong correlation and can be used both to construct model Hamiltonians and to perform full \textit{ab initio} calculations 
\cite{deGraaf2016,10.1021/jp013554c,cui2022}%
.
These methods are, however, computationally demanding and their cost grows steeply with system size. 

Thanks to the local nature of electronic correlations~\cite{doi:10.1142/8419} -- nearest-neighbour magnetic coupling $J_1$ is often determined by only the two interacting magnetic centres --, embedded-cluster approach offers an affordable \textit{ab initio} route for applying correlated quantum-chemistry methods to crystalline materials~\cite{Moreira1997,Illas1997,Su1999,Munoz2000,deGraaf2016}. Embedded-cluster quantum chemistry thus has a long record in the determination of exchange couplings in cuprates and nickelates: La$_2$CuO$_4$, Sr$_2$CuO$_3$, and NdNiO$_2$, for example, have been studied with
different methods with reasonable accuracy against experiment~\cite{Munoz2000,Calzado2001,Bogdanov2022,katukuri2020,katukuri2022,cui2022,jiang2025}. There are, however, fewer studies of how the magnetic coupling changes with the bridging angle, including the sign change from AFM to the FM phase \cite{doi:10.1021/ct100329n}; benchmarks of how robust \textit{ab initio} methods are in these regimes remain scarce. Furthermore, the nickelates that have been discovered relatively recently, have only a few studies with wave function methods~\cite{katukuri2020,katukuri2022}. 

It has been recently shown \cite{Bogdanov2022} that describing superexchange requires the inclusion of the so-called orbital breathing effects into the description of the system. The breathing emerges from the on-site virtual 3d--4d interaction of the magnetic sites,
effectively extending the 3d orbitals. This reduces the effective $U$ from about 28\,eV deduced from a Cu Coulomb integral to about 8-10\,eV in cuprates~\cite{olleA,olleB,olleC,olleD} and simultaneously increases the effective hopping between the Cu atoms~\cite{olleE}. These two effects are crucial in increasing the magnitude of $J$. In establishing these results, however, it was found necessary to use large active-space solvers, augmented with perturbation theory, which clearly has the disadvantage of being computationally expensive and difficult to scale to larger systems.  

The purpose of this work is two-fold. First, we refine our understanding of orbital-breathing enhanced correlation in these materials, and use the insight to demonstrate that computationally affordable 2-determinant coupled cluster \cite{balkovaCoupledcluster1992} (2det-CC) and its distinguishable cluster (2det-DC) variant \cite{schraivogelTwoDeterminant2024} (2det-CCSD and 2det-DCSD in this work, since we do not include excitations beyond doubles) are well suited for evaluating $J_1$ without having to resort to large active space solvers. With this method, we reach larger basis sets than -- to our knowledge -- used before with wave-function methods in these systems, allowing us to compute complete basis set limit (CBS) estimates of $J_1$. Our study shows that basis sets used in previous studies, where a low level of basis was considered sufficient~\cite{10.1021/jp013554c,10.1063/1.466769,10.1063/1.473065}, are at least with the coupled cluster method not enough to eliminate the finite-basis error. Second, we apply {\em transcorrelation (TC) theory}, for the first time, to these strongly correlated solids, demonstrating that accurate $J_1$'s can be obtained in relatively small basis sets, thanks to the drastic acceleration of basis-set convergence of the TC method.

In TC, the electronic Hamiltonian is similarity-transformed with respect to a real-space Jastrow factor, an interparticle correlator with up to two-electron terms. Although TC was suggested decades ago~\cite{boys1969}, it has only recently been made practical, thanks to sophisticated Jastrow optimization routines~\cite{haupt2023,filip2025jastrow}, downsampling of tedious $3$-body interaction terms via the xTC-approximation~\cite{christlmaier2023}, and implementation of a range of highly accurate second-quantized correlation solvers able to treat non-hermitian Hamiltonians~\cite{schraivogel2021,schraivogel2023,kats2024,luo2018combining,haupt2025,baiardi2022explicitly,liao2023dmrg,dobrautz2024toward}. 

The route to TC in the demanding setting of strongly-correlated oxides was established in three stages: transcorrelated theory with pseudopotentials~\cite{simula2025ecp}, its extension to transition-metal atoms and their spin excitations~\cite{simula2025tm}, and the transcorrelated treatment of periodic bulk and defect systems within embedding approaches~\cite{simula2026}. The present study combines all three and applies TC to embedded-cluster simulations of correlated transition-metal oxides, where it uniformly improves the accuracy of the 2det-CC methods in any single basis set, and closely reproduces their CBS estimates of $J_1$ already at the relatively small composite triple-$\zeta$ basis (see below for the definition of composite bases).

We start with a simplified model with breathing terms, taken from Ref.~\cite{Bogdanov2022}. Comparing full diagonalization results with those of Rayleigh-Schr\"{o}dinger perturbation theory (RSPT) and coupled cluster (CC), we explain how breathing and hopping couple in a correct treatment, and how methods more affordable than full configuration interaction (FCI) within a large active space can still capture the important interactions. Then we show that the established methods, combined with TC Hamiltonians, determine $J_1$ with high accuracy in realistic cuprate materials across a variety of bridging-oxygen bond angles, including the sign flip of $J_1$ from AFM to FM when we model a material with $94^\circ$ Cu-O-Cu bridging angle. For NdNiO$_2$, we establish more robust (Hubbard-constrained) derivation of $J_1$ from published experimental data~\cite{lu2021}, and show that it closely agrees with our computed result, much better than the originally reported value. We also move beyond the $2$-magnetic-site embedding model and compute the nearest-neighbour coupling using $4$-magnetic-site clusters in La$_2$CuO$_4$ and NdNiO$_2$. A close agreement of our $J_1$ obtained for La$_2$CuO$_4$ with earlier coupled cluster calculations with more advanced embedding~\cite{cui2022} in low-level basis connects our calculations to previous theoretical studies. This work thus establishes TC as a mature theory able to reliably tackle problems of strongly correlated materials.  

\section{Capturing Orbital Breathing}

In recent study~\cite{Bogdanov2022} the orbital breathing was described as the virtual 3d-4d orbital interaction occurring in an intermediate Cu 3d$^{10}$ state, that lowers the effective Cu 3d-3d Coulomb repulsion and increases the effective hopping between a Cu 3d orbital and an oxygen 2p orbital. Multireference second-order perturbation theory approaches, namely complete active space perturbation theory (CASPT2) and $n$-electron valence state perturbation theory (NEVPT2), therefore fail to provide accurate values of $J_1$ unless all of the Cu 3d and 4d and O 2p and 3p orbitals are included in the active space that is solved with a method of FCI accuracy. 

The effect of breathing to the on-site Coulomb interaction and hopping was demonstrated with a simplified model system, consisting of two sites that represent the magnetic transition metal ion pair, and two orbitals on both sites, representing the 3d and 4d orbitals. The Hamiltonian was defined with intersite hoppings $t_{ij}$, on-site Coulomb interaction $U_{ij}$, and on-site breathing terms $K_i$, together with orbital energies $\varepsilon_i$. The indices $i$ and $j$ take value 1 for the $3$d and value 2 for the $4$d orbitals of the two Cu sites. The analysis was restricted to three states, each a configuration state function. We provide a re-derivation of the model Hamiltonian for the singlet and triplet states in the Supplemental Material~\cite{Supplement}.

To demonstrate that the breathing effects can also be captured by methods of lower level than FCI, we solved the singlet and triplet blocks of this model Hamiltonian -- a $3\times3$ and a $2\times2$ matrix respectively -- by exact diagonalization, Rayleigh--Schr\"odinger perturbation theory, and coupled cluster. We perform two variants of RSPT: with diagonal and breathing references. The latter removes the $K_1$ terms from the perturbation and includes them into the reference Hamiltonian, that is solved with exact diagonalization. Hence the RSPT with breathing reference corresponds to a model equivalent of multireference perturbation theory, such as CASPT2 or NEVPT2. The theory and algorithms for the four methods are given in a Jupyter notebook included in the Supplemental Material~\cite{Supplement}, which can reconstruct the results summarized in \Cref{tab:model_j}.

With all methods, we reproduce the results of Ref.~\cite{Bogdanov2022}. Exact diagonalization amounts to an FCI treatment. In diagonal RSPT, the third order of perturbation adds a correlation energy term $8\,t_{11}t_{12}K_1/(U_{11}\Delta)$, with $\Delta = \varepsilon_2 - \varepsilon_1 + U_{12}$, in which hopping and breathing are coupled; $t_{11}$ is the intersite hopping between the two lower (3$d$) orbitals, $t_{12}$ the intersite hopping from a lower to a higher (4$d$) orbital, and $U_{11}$ the on-site Coulomb repulsion, see the Supplemental Material~\cite{Supplement}. This order offers the largest improvement in $J_1$, but orders at least up to order $8$, and thus more complex couplings of breathing and hopping, are required to evaluate $J_1$ correctly. The breathing reference RSPT captures the correct value of $J_1$ already at second-order level. This explains the success of CASPT2 and NEVPT2 approaches with large active spaces. With $K_1=0$, the second order of perturbation is already practically converged. The synergetic effect of breathing and charge transfer is thus captured only by high-order interaction patterns in which breathing and charge transfer (inter-site hopping) are coupled. 

\begin{table}[!htb]
\centering
\caption{Singlet--triplet gap $E_\text{ST} = E_T - E_S$ (meV) of the two-site breathing model, computed with the model parameters of Ref.~\cite{Bogdanov2022}, with the breathing coupling active ($K_1\!=\!-8$~eV) and switched off ($K_1\!=\!0$). Exact diagonalization, Rayleigh--Schr\"odinger perturbation theory (RSPT) at increasing order, and coupled cluster all use the same $3\times3$ singlet and $2\times2$ triplet matrices. The breathing coupling enhances the gap by a factor $J_1(K_1)/J_1(0) = 2.25$.}
\label{tab:model_j}
\small
\begin{tabular}{@{}lcc@{}}
\toprule
Method & $K_1 = -8$~eV & $K_1 = 0$ \\
\midrule
Exact diagonalization & 171.73 & 76.29 \\
\midrule
\multicolumn{3}{@{}l}{\textit{RSPT ($K_1$ in the perturbation)}} \\
\quad order 2  &  76.92 & 76.92 \\
\quad order 3  & 120.90 & 76.92 \\
\quad order 4  & 147.94 & 76.29 \\
\quad order 6  & 166.27 & 76.29 \\
\quad order 8  & 170.62 & 76.29 \\
\quad order 10 & 171.55 & 76.29 \\
\quad order 14 & 171.73 & 76.29 \\
\midrule
\multicolumn{3}{@{}l}{\textit{RSPT ($K_1$ in the reference)}} \\
\quad order 2  & 176.19 & 76.92 \\
\quad order 3  & 176.19 & 76.92 \\
\quad order 4  & 171.52 & 76.29 \\
\quad order 6  & 171.74 & 76.29 \\
\quad order 8  & 171.73 & 76.29 \\
\midrule
Coupled cluster & 171.73 & 76.29 \\
\bottomrule
\end{tabular}
\end{table}

Coupled cluster provides the $J_1$ corresponding to that of the exact diagonalization. This is expected, since the excitation operator for the CC \textit{ansatz} contains all the possible excitations for the simple model system, and thus coupled cluster is exact by definition. What carries over to a real material is the structure of the problem: second-order perturbation theory is insufficient by construction, but CC methods can couple excitation and breathing processes. Moreover, while coupled cluster in general is known to be inaccurate for strongly correlated or multireference systems, the $2$-determinant variant we use works -- as the CC algorithm built for the model system in the Supplement~\cite{Supplement} -- with configuration state functions instead of Slater determinants as reference state representations. Therefore for the nearest-neighbour magnetic coupling 2det-DCSD is likely a very useful tool that directly targets the correct strongly correlated states, whereas with single-determinant coupled cluster we would have to rely on broken-symmetry references and incomplete representation of the singlet states. 

Based on the above considerations we set out to test if 2-determinant variant of coupled cluster is able to capture the breathing effects that are crucial for accurate determination of the nearest-neighbour magnetic parameters, and will now proceed to test its accuracy in real, challenging cases.

\section{Superexchange in real materials}

\subsection{Materials and Embedded Clusters}

\begin{figure*}[t]
    \centering
    \includegraphics[width=\textwidth]{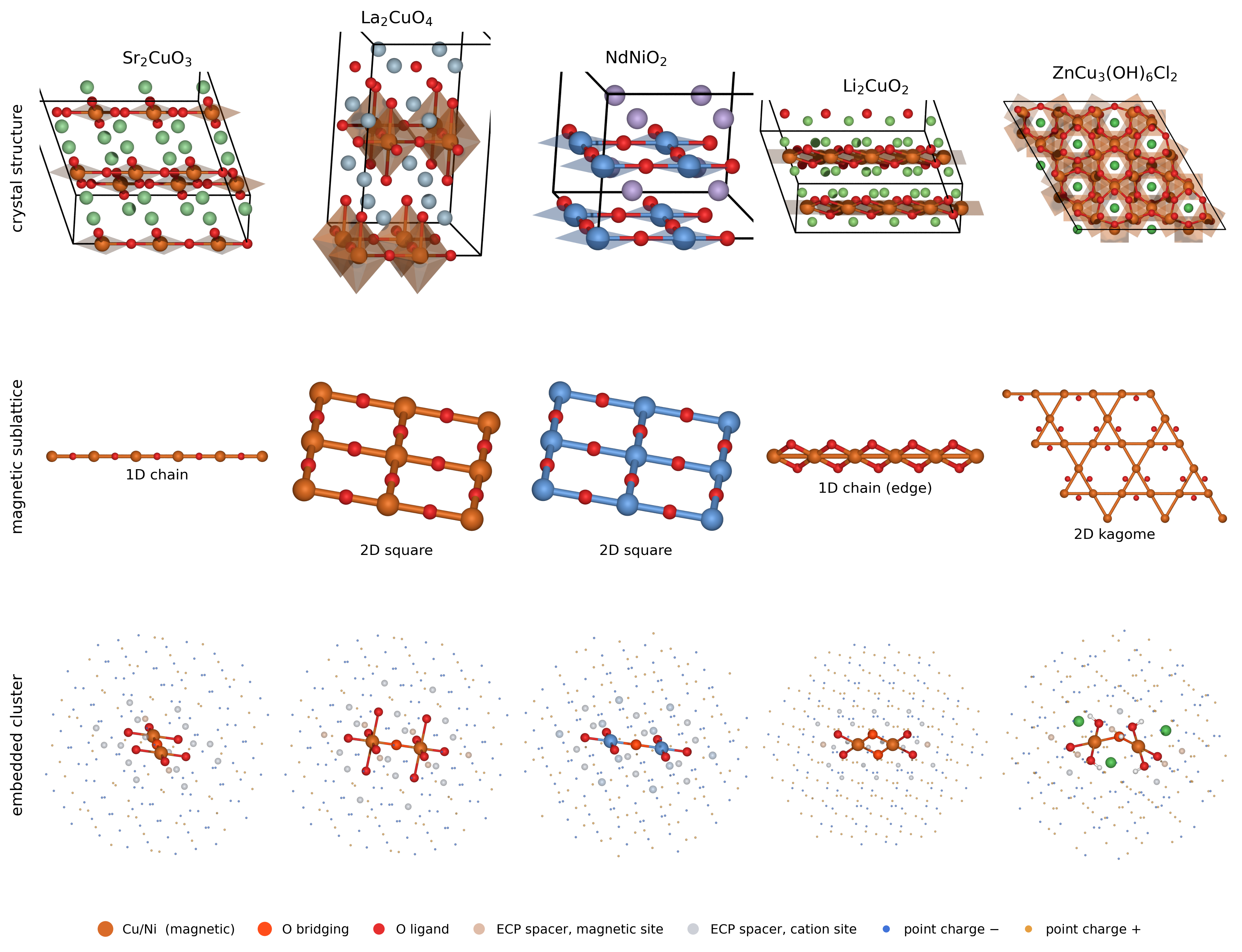}
    \caption{Top row: the structures of the materials studied in this work. Middle row: the magnetic sublattices of the materials. Bottom row: the atomic embedding clusters used in the calculations. The embedding point-charge clouds at the atomic sites of the materials are depicted only in the proximity of the embedding clusters.  Figure prepared with AI assistance and verified by the authors; see Acknowledgments.}
    \label{fig:structures}
\end{figure*}

\begin{figure}[t]
    \centering
    \includegraphics[width=\columnwidth]{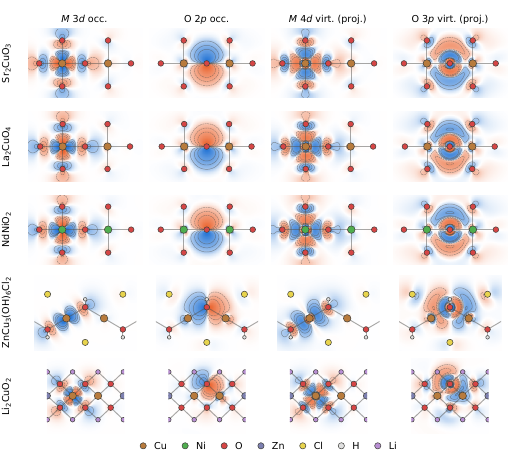}
    \caption{Localized CASSCF orbitals in the magnetic-sublattice plane of
    each material: the magnetic $3d$ and bridging-oxygen $2p$ occupied
    orbitals, and their virtual-space breathing partners, obtained by
    projecting the correspondingly oriented $4d$ ($3p$) atomic orbital onto
    the virtual space. Blue/orange: positive/negative amplitude on a
    symmetric log scale; contour lines at $\pm 0.02$, $0.05$, $0.1$~a.u.  Figure prepared with AI assistance and verified by the authors; see Acknowledgments.}
    \label{fig:orbitals}
\end{figure}

We modeled the magnetic interactions in Sr$_2$CuO$_3$, La$_2$CuO$_4$, Li$_2$CuO$_2$, NdNiO$_2$, and ZnCu$_3$(OH)$_6$Cl$_2$ (herbertsmithite) by using experimental structures~\cite{hyatt2004,radaelli1994,sapina1990,hayward2003,shores2005}. The periodic unit cells of the materials are shown in the top row of Fig.~\ref{fig:structures}. Their magnetic sublattices are represented in the middle row. This set of materials represents a range of different chemical environments. Each material has a different spacer atom, creating a spacer set of Sr, La, Nd, Li, and Zn. Magnetic sublattices are 1D corner-sharing chain, 2D lattice, 1D edge-sharing chain, and Kagome lattice; the two presented 2D sublattices differ by the magnetic-site ions, which leads to substantial differences in electronic correlations \cite{katukuri2020,katukuri2022}, plus La$_2$CuO$_4$ has apical oxygens. The 1D CuO$_2$ chains in Li$_2$CuO$_2$ are ferromagnetic (FM) because of the hopping-suppressing Cu-O-Cu angle of $94^\circ$, while rest of the magnetic sublattices are antiferromagnetic. Therefore the presented set establishes a demanding test for the ability of the coupled cluster, embedding model, and transcorrelated methods to capture the chemical ligand effects and the physics of the magnetic properties of the materials.

Following the embedded-cluster approach \cite{deGraaf2016, chargedel}, we constructed finite clusters containing two active Cu or Ni sites, their neighbouring ligands, and spacer atoms without electrons or basis functions; the basis sets, pseudopotentials, and embedding parameters of each system are collected in Appendix B. Clusters used in the calculations, together with the surrounding point-charge embedding clouds (but distant point charges omitted for clarity) that reproduce the Madelung potential due to the infinite lattice, are illustrated in the bottom row of Fig.~\ref{fig:structures}. 

In the minimal active space of the embedded clusters, containing only the Cu $3d_{x^2-y^2}$ orbitals, we optimized the CASSCF(2o,2e) orbitals by taking state-averaged solutions over singlet and triplet states, and localized the orbitals (Appendix B gives a more detailed description of orbital generation). Out of these orbitals we constructed non-TC second-quantized Hamiltonians. To compute the transcorrelated corrections to this Hamiltonian, we optimized the Jastrow factors within the point-charge cloud, using the 2-determinant singlet state as the trial wave function (the similarity transformation preserves the eigenenergies of the Hamiltonian, and thus this Jastrow should reproduce also the triplet energy correctly), and evaluated the xTC-PP (transcorrelation with xTC approximation and pseudopotentials) corrections to the second-quantized Hamiltonian. See Appendix A for the xTC-PP theory. 

The procedure for generating Hamiltonians was repeated for a hierarchy of gaussian basis sets: for non-transcorrelated calculations, we used cc-pVDZ (DZ) basis for all atoms, with ccECP pseudopotentials for the active atoms. Then we substituted the basis of the two magnetic cations and the bridging oxygen(s) by cc-pVTZ (TZ/DZ) and cc-pVQZ (QZ/DZ) bases. We also did a calculation with cc-pVTZ for all atoms in the cluster (TZ), to get an estimate of the missed correlations of the outer ligands due to the use of composite bases. With transcorrelation, we used the recently established singular value decomposition-based virtual space compression~\cite{cheng2026compression}, where the reference orbitals and the Jastrow are obtained with all atoms in cc-pVQZ basis (QZ), but the virtual space is compressed to sizes of DZ, TZ/DZ, and QZ/DZ. The details of these steps are given in Appendices B and C; the average computational cost of each workflow step for the embedded clusters is reported in Appendix D (\Cref{tab:timings}).

Figure \ref{fig:orbitals} presents the Cu 3d$_{x^2-y^2}$ and O 2p occupied orbitals for each system, solved with the aforementioned procedure in the TZ/DZ basis. Also the virtual Cu 4d$_{x^2-y^2}$ and O 3p orbitals are presented, obtained by projecting the atomic orbitals of the Cu and O atoms into the virtual space. These orbitals, along with their degenerate counterparts, should all be included into the active space of a CASPT2 or NEVPT2 calculation, in order to get accurate magnetic couplings~\cite{Bogdanov2022}. However, we claim that coupled cluster methods can couple these orbitals up to a sufficient degree to capture the important correlations. We can see in Fig.~\ref{fig:orbitals} that the systems with a $180^\circ$ bridging angle have very similar and highly symmetric orbitals, but as the angle starts to bend towards $90^\circ$, this is no longer the case. In these bent angles the correlated solver should be able to balance between relevant correlations of the triplet and the singlet that arise in this orbital set in order to capture magnetic couplings correctly. 

\begin{figure}[t]
    \centering
    \includegraphics[width=\columnwidth]{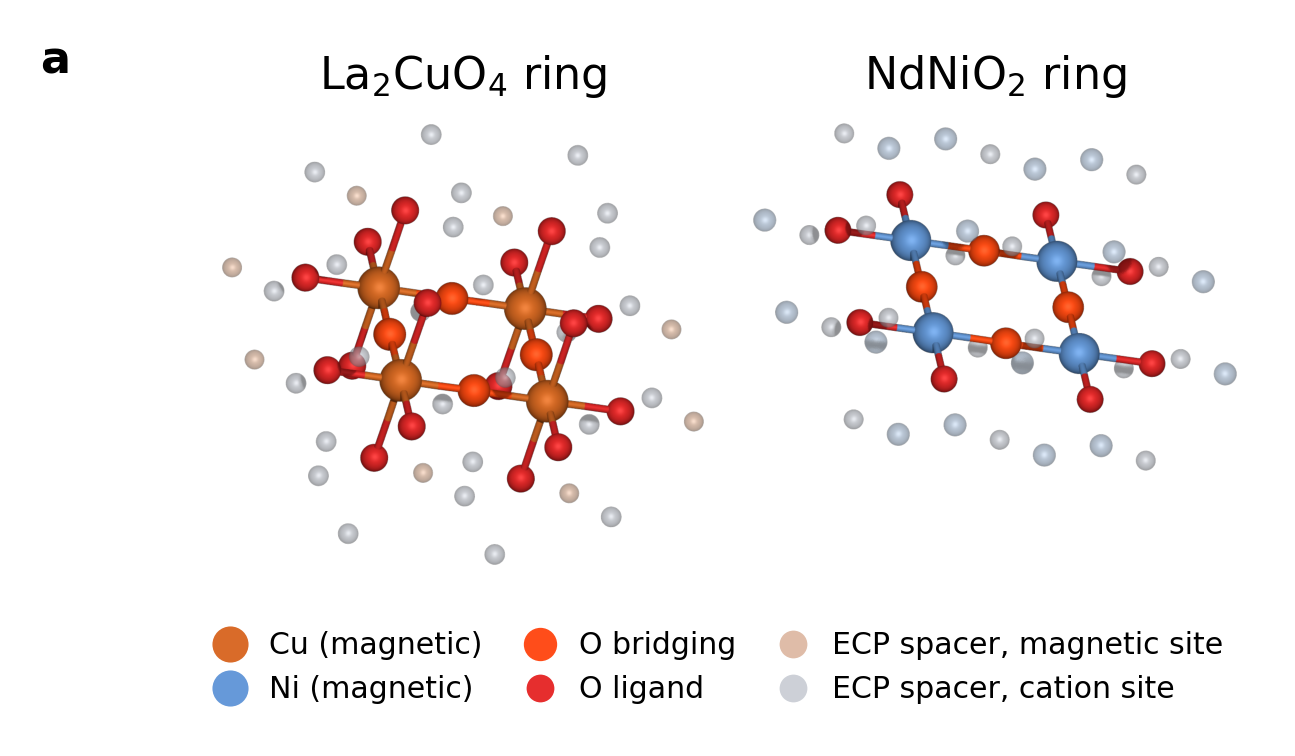}\\[3pt]
    \includegraphics[width=\columnwidth]{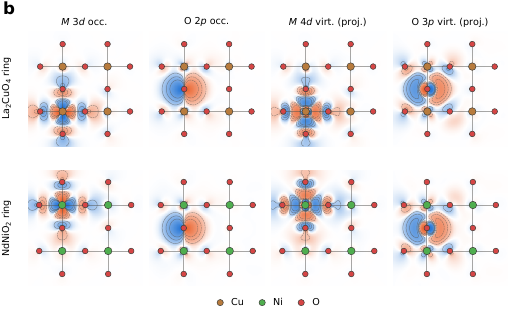}
    \caption{\textbf{a},~The $4$-site (ring) embedding clusters of
    La$_2$CuO$_4$ and NdNiO$_2$, in the same representation as the embedded
    clusters of Fig.~\ref{fig:structures}; the point-charge clouds are
    omitted. \textbf{b},~Localized CASSCF orbitals in the magnetic-sublattice
    plane of the $4$-site clusters, obtained with the same procedure as in
    Fig.~\ref{fig:orbitals}. Figure prepared with AI assistance and verified by the authors; see Acknowledgments.}
    \label{fig:4siteorbitals}
\end{figure}

To test the finite-size error in magnetic couplings of our 2-site embedding clusters, we built 4-site clusters for La$_2$CuO$_4$ and NdNiO$_2$. The clusters were built with an equal procedure, only by expanding the cluster with an additional 2-site pair of magnetic sites. The $4$-site clusters and their localized orbitals are depicted in Fig.~\ref{fig:4siteorbitals}.

\subsection{Comparison to Experimental Heisenberg Parameters}

Experiments adopt different strategies in estimating parameters of the Heisenberg Hamiltonian, based on the underlying magnetic sublattice, its geometry and the theoretical assumptions on the physics of the systems. We will list briefly the experimental methods, with more focus on the magnetic square lattices. 

In 2D magnetic sublattice, the Heisenberg Hamiltonian with up to four-spin terms can be written as 
\begin{align}
    \begin{aligned}
        \hat{H}&=E_0+J_1\sum_{\langle i,j\rangle}\vS_i\cdot \vS_j+J_2\sum_{\langle\langle i,j\rangle\rangle}\vS_i\cdot \vS_j\\&+J_3\sum_{\langle\langle\langle i,j\rangle\rangle\rangle}\vS_i\cdot \vS_j+J_c\sum_{\langle i,j,k,l \rangle}\hat{\mathbf{P}}, \\
        \hat{\mathbf{P}}&=(\vS_i\!\cdot\!\vS_j)(\vS_k\!\cdot\!\vS_l)
       + (\vS_i\!\cdot\!\vS_l)(\vS_j\!\cdot\!\vS_k)
       - (\vS_i\!\cdot\!\vS_k)(\vS_j\!\cdot\!\vS_l),
    \end{aligned}
\label{eq:heisenenberghamiltonian}
\end{align}
where the sum $\langle i,j\rangle$ runs over nearest-neighbour $i$ and $j$ pairs, $\langle\langle i,j\rangle\rangle$ ($\langle\langle\langle i,j\rangle\rangle\rangle$) over second-nearest diagonal pairs (third-nearest pairs, second-nearest along a Cu--O--Cu chain), and $\langle i,j,k,l\rangle$ over neighbours within $4$-site plaquettes like the ones in Fig.~\ref{fig:4siteorbitals}.

Coldea \textit{et al.} measured inelastic neutron scattering (INS) data of spin wave dispersion over the Brillouin zone~\cite{coldea2001} in La$_2$CuO$_4$. Using linear spin wave theory~\cite{spinw}, they fitted the magnetic parameters in Eq.~\ref{eq:heisenenberghamiltonian} by comparing the parameter-based dispersion curve with the INS data.

First Coldea \textit{et al.} fitted a Heisenberg model with only nearest- and second-nearest neighbour spin couplings, $J_1$ and $J_2$, non-zero, giving what we from now on call the effective spin couplings $J_\text{eff}$ and $J^{'}_\text{eff}$. At $10$~K, they then got $J_\text{eff}=104.1 \pm 4$~meV and $J^{'}_\text{eff}=-18\pm 3$~meV. Because they found a ferromagnetic $J'_\text{eff}$ to be against theoretical predictions, they did another fit with the full Hamiltonian in Eq.~\ref{eq:heisenenberghamiltonian}, but by restricting the magnetic coupling parameters to obey the single-band Hubbard model, with nearest-neighbour intersite hopping $t$ and  Coulomb on-site repulsion $U$ as parameters: $J_1=4t^2/U-24t^4/U^3$, $J_c=80t^4/U^3$ and $J_2=J_3=4t^4/U^3=J_c/20$. This gave an antiferromagnetic $J_2=3.05\pm 0.4$~meV and a $J_1=146.3\pm 4$~meV. The findings were confirmed in a later study~\cite{Headings2010} with $J_1=143\pm2$~meV. The relation between the Hubbard-constrained $J_1$ and effective $J_\text{eff}$ and $J^{'}_\text{eff}$ can be also drawn.  This relation follows from requiring the Hubbard-constrained model to reproduce the two-parameter magnon energies at reciprocal points $(\pi,0)$ and $(\pi/2,\pi/2)$. For La$_2$CuO$_4$, the two fits of Coldea \textit{et al.} are related by $J_1 = J_\text{eff} - \tfrac{7}{3}J'_\text{eff}$ to within 0.2~meV (146.1 against 146.3~meV at 10~K, 138.4 against 138.3~meV at 295~K), see Supplemental Material for derivation~\cite{Supplement}.

Lu \textit{et al.}~\cite{lu2021} measured with resonant inelastic X-ray scattering (RIXS) the spin wave dispersion in $10$~nm thick NdNiO$_2$ sample, and fitted the Heisenberg model for $J_\text{eff}$ and $J^{'}_\text{eff}$, yielding values $J_\text{eff}=63.6\pm 3.3$~meV and $J^{'}_\text{eff}=-10.3\pm 2.3$~meV. The authors discuss the ferromagnetic $J^{'}_\text{eff}$ and reasons behind it, suggesting possible long-range metallic exchange mediated by the Nd $5$d pockets. However, if we apply the relation between $J_1$ and the effective parameters in the way that worked accurately in La$_2$CuO$_4$, we get $J_1 = 87.6$~meV for NdNiO$_2$. We will present both values in the section with our results for $J_1$.

For Sr$_2$CuO$_3$, the solution to the 1D Heisenberg chain is exact, allowing direct fits to exact two- and four-spinon structure factors from RIXS~\cite{Schlappa2012} and INS data~\cite{Walters2009}. These two experiments report values of $J_1=249$~meV and $J_1=241\pm11$~meV, respectively. For Li$_2$CuO$_2$, magnetic dispersion was measured with INS, and results fitted to Heisenberg Hamiltonian with linear spin wave theory. Zoghlin \textit{et al.} obtained $J_1=-18.4\pm0.08$~meV at $T=5$~K~\cite{zoghlin2023}. Another value, $J_1=-19.7$~meV, was measured for Li$_2$CuO$_2$ by Lorenz \textit{et al.}~\cite{Lorenz_2009}. The experimental $J_1=14.7$~meV for herbertsmithite was estimated in Ref.~\cite{rigol2007} by fitting the temperature dependence of the magnetic susceptibility to temperature. 

\subsection{Magnetic coupling from \textit{ab initio}}

In 2-site embedding clusters we estimate the magnetic coupling as $J_1=E_T-E_S$, where $E_T$ and $E_S $ are 2-determinant coupled cluster energies of the triplet and singlet states of the clusters, respectively. The reference states are, in terms of Slater determinants,  $\Psi_T=1/\sqrt{2}\left(|\phi_A^\uparrow\phi_B^\downarrow|+|\phi_A^\downarrow\phi_B^\uparrow|\right)$ and $\Psi_S=1/\sqrt{2}\left(|\phi_A^\uparrow\phi_B^\downarrow|-|\phi_A^\downarrow\phi_B^\uparrow|\right)$, where orbitals $\phi_L^\sigma$ are the localized 3d$_{x^2-y^2}$ orbitals of sites $L=A,B$, of spin $\sigma$, and the Slater determinants are expressed without explicitly writing down the doubly occupied orbitals. An alternative way of estimating $J_1$ is with single-reference CC, by computing energies of the $m_s=1$ triplet and the broken-symmetry (BS) $m_s=0$ states, defined as $\Psi_\text{FM}=|\phi_A^\uparrow\phi_B^\uparrow|$ and $\Psi_\text{BS}=|\phi_A^\uparrow\phi_B^\downarrow|$, respectively. The $J_1$ is then estimated as $J_1=2(E_\text{FM}-E_\text{BS})$. These \textit{ab initio} approaches offer access to nearest-neighbour magnetic coupling based only on the Hamiltonian we build under the embedding conditions, and offer an estimate independent of assumptions on the form of the Heisenberg Hamiltonian or the Hubbard-like constraints for the coupling parameters.

We compute $J_1$ in the $4$-site fragments for La$_2$CuO$_4$ and NdNiO$_2$  by estimating single-reference CC energies for two reference states: FM and AFM (N\'eel), defined as 
\begin{align}
  \begin{aligned}
    \Psi_\text{FM}&=|\phi_A^\uparrow\phi_B^\uparrow\phi_C^\uparrow\phi_D^\uparrow|\\
    \Psi_\text{AFM}&=|\phi_A^\uparrow\phi_B^\downarrow\phi_C^\uparrow\phi_D^\downarrow|\\
  \end{aligned}
\end{align}
where sites A, B, C, and D label the 4 magnetic sites in a clockwise order. Then we determine $J_1$ as 
\begin{align}
\begin{aligned}
  J_1&=\tfrac12\!\left(E_{\mathrm{FM}}-E_{\mathrm{AFM}}\right).
\label{eq:inversion}
\end{aligned}
\end{align}
In this calculation, the $J_c$ is present in the model, but because its contribution to FM and AFM energies is identical, it cancels out in the expression of $J_1$.

\subsection{Nearest-neighbour $J_1$ Values with 2-determinant Coupled Cluster}

\begin{figure*}[!t]
    \centering
    \includegraphics[width=\textwidth]{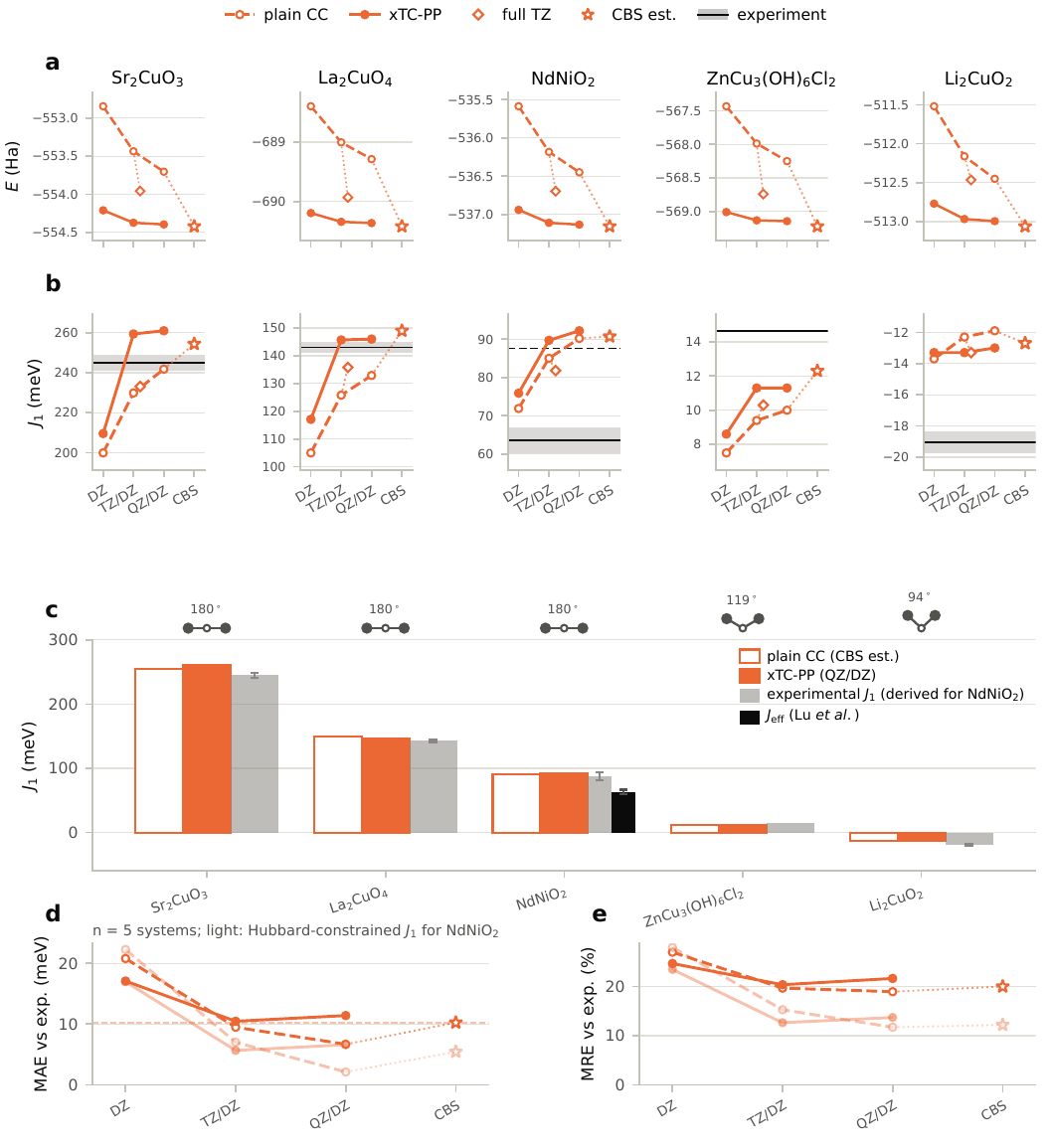}
    \caption{Total energies and spin-coupling constant $J_1$ from 2det-DCSD, without (non-TC CC: dashed lines, open symbols) and with the xTC-PP treatment (solid lines, filled symbols). \textbf{a},~Total energy of the singlet state: non-TC CC over the composite bases, with the full-TZ energy as the offset diamond beside TZ/DZ, continued to its CBS estimate (star), and xTC-PP at the DZ, TZ/DZ and QZ/DZ projections, as computed. \textbf{b},~Basis-set convergence of $J_1$, with method and basis notation as in (a); the horizontal black line is the experimental value and the grey band its reported uncertainty; for NdNiO$_2$ the black dashed line is the Hubbard-constrained $J_1$ (see Supplement~\cite{Supplement}). \textbf{c},~2det-DCSD against experiment~\cite{Schlappa2012,Walters2009,Headings2010,Lorenz_2009,zoghlin2023,lu2021,rigol2007} (grey bars), systems ordered by descending $J_1$: the open bar is the non-TC-CC CBS estimate and the filled bar the TC value at QZ/DZ; for NdNiO$_2$ the grey bar is the Hubbard-constrained $J_1$ derived from the fitted $J_\mathrm{eff}$ of Lu \textit{et al.}~\cite{lu2021} (black bar); the nearest-neighbour M--O--M bridging angle (M = Cu, Ni) of each system is sketched above its bars. \textbf{d},\,\textbf{e},~Mean absolute (\textbf{d}) and mean relative (\textbf{e}) error with respect to experiment, averaged over the systems with an experimental reference and results from both variants (dark), and with the Hubbard-constrained $J_1$ as the reference for NdNiO$_2$ (light); the basis axis and symbols follow the notation of (a). Figure prepared with AI assistance and verified by the authors; see Acknowledgments.}
    \label{fig:2det_comparison}
\end{figure*}

With the constructed Hamiltonians of the 2-site embedding clusters, we performed the 2det-CCSD and 2det-DCSD calculations. The prefix xTC-PP on a method name indicates that the 2det-CC calculation was performed with the xTC-PP Hamiltonian of Eq.~\eqref{eq:xtc_hamiltonian}. The results are shown in \Cref{tab:2det_main} and Fig.~\ref{fig:2det_comparison}, where for clarity, we only present 2det-DCSD results, which we found to be more accurate than 2det-CCSD, see Supplementary material~\cite{Supplement}. 

\begin{table*}[!t]
\centering
\caption{Singlet--triplet coupling constant $J_1$ (meV) from 2-determinant methods. TC = transcorrelated (xTC-PP). The non-TC-CC values are the CBS estimates. The TC values are at the TZ/DZ and QZ/DZ correlation bases with the QZ Jastrow factor. The corresponding 2det-CCSD results are given in Table~S1 of the Supplementary material~\cite{Supplement}. The theory references are CASSCF+NEVPT2~\cite{Bogdanov2022} for \ch{Sr2CuO3} and \ch{La2CuO4}, and CASSCF-MR-LCC~\cite{katukuri2020} for \ch{NdNiO2}}
\label{tab:2det_main}
\small
\sisetup{table-format=-3.1}
\setlength{\tabcolsep}{7pt}
\begin{tabular}{@{}l SSS @{\hspace{14pt}} cc@{}}
\toprule
 & {2det-DCSD} & \multicolumn{2}{c}{2det-xTC-PP-DCSD} & \multicolumn{2}{c}{Reference} \\
\cmidrule(lr){2-2}\cmidrule(lr){3-4}\cmidrule(l){5-6}
System & {CBS} & {TZ/DZ} & {QZ/DZ} & Theory & Experiment \\
\midrule
\ch{Sr2CuO3}           & 254.4 & 259.4 & 261.0 & 247~\cite{Bogdanov2022} & 241--249~\cite{Schlappa2012,Walters2009} \\
\ch{La2CuO4}           & 149.0 & 145.7 & 146.0 & 137~\cite{Bogdanov2022} & $143 \pm 2$~\cite{Headings2010} \\
\ch{NdNiO2}            &  90.7 &  89.7 &  92.2 & 78~\cite{katukuri2020}  & $63.6 \pm 3.3$~\cite{lu2021} ($87.6^{*}$) \\
ZnCu$_3$(OH)$_6$Cl$_2$ &  12.3 &  11.3 &  11.3 & ---                    & 14.65~\cite{rigol2007} \\
\ch{Li2CuO2}           & -12.7 & -13.3 & -13.0 & ---                    & $-19.7$~\cite{Lorenz_2009}, $-18.4$~\cite{zoghlin2023} \\
\bottomrule
\end{tabular}

\vspace{3pt}
{\footnotesize $^{*}$Hubbard-constrained $J_1$, derived from the fit of Lu \textit{et al.}~\cite{lu2021}; see Supplement~\cite{Supplement}.}
\end{table*}

In Fig.~\ref{fig:2det_comparison}, panel (a), we present the total 2det-DCSD energy of the singlet state; the triplet energy coincides with it on this scale. In non-TC calculations we present also a CBS estimate based on extrapolation to infinite composite basis limit, and a correction based on the TZ and TZ/DZ energies~(see Appendix B on how CBS energies were estimated). The xTC-PP energies are much closer to total energy CBS estimates already at the DZ projection than the non-TC variant, and in TZ/DZ and QZ/DZ composite bases land very close to the non-TC-CC CBS estimates. 

In panel (b), we present the basis-set convergence of the magnetic coupling $J_1$, as extracted from the total energies. The results are compared against experimental estimates for Sr$_2$CuO$_3$~\cite{Schlappa2012,Walters2009}, La$_2$CuO$_4$~\cite{Headings2010}, Li$_2$CuO$_2$~\cite{Lorenz_2009,zoghlin2023}, NdNiO$_2$~\cite{lu2021} (the value of the reference is $J_\text{eff}$; thus we also present the derived $J_1$, see Supplementary material~\cite{Supplement}), and herbertsmithite \cite{rigol2007}. The basis-dependence of the non-TC results is revealing: the difference of DZ and CBS results can be as large as 50 meV. The QZ/DZ basis does not seem to provide full convergence, and using TZ instead of composite TZ/DZ provides a meaningful correction, which is positive for Sr$_2$CuO$_3$, La$_2$CuO$_4$ and herbertsmithite, and negative for Li$_2$CuO$_2$ and NdNiO$_2$. These results show that the basis sets usually studied and considered convergent~\cite{10.1063/1.473065,10.1063/1.466769,10.1021/jp013554c,Bogdanov2022,jiang2025} still contain error, at least with coupled cluster. In Supplement~\cite{Supplement} we compare $J_1$ from previous studies with multireference methods such as NEVPT2 and CASPT2  to our results and show that they provide, when comparing in matching basis set sizes, higher values than ours. Assuming that the finite-size effects are the same, this would mean those small-basis reference values are overestimations, but the basis set error of coupled cluster as opposed to active space FCI and perturbative methods can behave differently.

The xTC-PP results for $J_1$ are converged within a few meV already at TZ/DZ virtual space basis (reference energy is always full QZ quality), and match very closely to non-TC CBS estimates. This is similar to the convergence of the total energy in panel (a). Hence the use of TC with the SVD-based virtual space compression drastically reduces the basis set requirements for accurate $J_1$, allowing a CBS-quality estimation within a single, affordable basis set.

In panel (c), we show the non-TC CBS estimate, and the xTC-PP QZ/DZ estimate, of each material against the experimental value. The theoretical results closely follow the experimental $J_1$, even at the very small positive value of herbertsmithite, and correctly describe the change from antiferromagnetic to ferromagnetic coupling. In panels (d) and (e), we show the absolute and relative errors of the non-TC and xTC-PP methods in different basis sets, and for the CBS estimate. Again, for the TZ/DZ and QZ/DZ bases, the errors of the xTC-PP method very closely match those of non-TC CBS results.

Table~\ref{tab:2det_main} presents the CBS (non-TC) and TZ/DZ and QZ/DZ (xTC-PP) results, against experiment, numerically for each material. Also the theoretical references are presented, NEVPT2~\cite{Bogdanov2022} for Sr$_2$CuO$_3$ (TZ/DZ) and La$_2$CuO$_4$ (DZ), and MR-LCC2~\cite{katukuri2020} for NdNiO$_2$.

The average errors in panels (d) and (e) are shown with both $J_\text{eff}$ and $J_1$ used as reference for NdNiO$_2$. The appropriate reference is $J_1$: it gives the nearest-neighbour coupling of the full Hamiltonian of Eq.~(\ref{eq:heisenenberghamiltonian}), as does the constrained fit used for La$_2$CuO$_4$, whereas $J_\text{eff}$ is a parameter of the two-parameter model. Our 2-site values of 89.7--92.2~meV agree with the derived $J_1 = 87.6\pm6.3$~meV; they are 41--45\% above $J_\text{eff}=63.6\pm3.3$~meV. The same relation gives $J_c=34$~meV and $J_2=J_3=1.7$~meV. The ferromagnetic $J'_\text{eff}$ from the experimental 2-parameter fit is then compensating the absence of ring exchange, as in La$_2$CuO$_4$, and no long-range metallic exchange is needed to explain it. However, our derivation for the experimental, Hubbard-constrained $J_1$ assumes that the single-band Hubbard relations hold in NdNiO$_2$.

The signed relative deviations reveal that the relative error in the cuprates correlates with the bridging angle (sketched above the bars in panel (c)): For the three $180^\circ$ systems, where the breathing enhancement is largest, our $J_1$ agrees with experiment to within 2--7\%. As discussed in the Introduction, when the angle is reduced, the superexchange channel is gradually diminished and another exchange channel via two orthogonal p-orbitals of the bridging oxygens arises for the triplet. This may lie behind the larger relative errors for herbertsmithite and Li$_2$CuO$_2$ against experiment, but the small magnitude of $J_1$ in these materials can also cause large relative errors with absolute errors being similar with the $180^\circ$ systems.

The agreement for the three $180^\circ$ systems shows that coupled cluster captures the breathing-enhanced correlations; without them the results would be qualitatively wrong~\cite{Bogdanov2022}. The 2-determinant coupled cluster is also a balanced method for singlet-triplet gaps and particularly suited for this purpose here. However, generalization of coupled cluster to beyond two-determinant references is possibly needed for the study of more complex magnetic correlations. 

\subsection{Finite-size tests with 4-magnetic-site clusters}

The possible errors that can still lie in our simulations are likely due to the finite ionic embedding or correlation solver.
Cui \textit{et al.}~\cite{cui2022} studied the finite size effects with CC methods and density matrix embedding theory (DMET), and found $2$-magnetic-site fragments to be quite close to being convergent with their self-consistent density matrix embedding theory approach, when tested in CaCuO$_2$. The test was done in small basis sets of cc-pVDZ or smaller size. The same study used density matrix renormalization group methods to verify that CCSD is a sufficient correlation solver, although also in small basis set. These tests would suggest that there are only small errors in our results too, but leave some room for uncertainty, as errors in finite size, basis, and correlation may be correlated. 

\begin{table}[h]
\centering
\caption{Comparison between computed $J_1$ obtained with ionic embedding with ECPs, and with DMET embedding without ECPs. Both estimates used CCSD in a $4$-site cluster according to Eq.~\ref{eq:inversion} using basis sets of double-$\zeta$ quality.}
\label{tab:cui comparison}
\footnotesize
\setlength{\tabcolsep}{3.5pt}
\begin{tabular}{@{}lcr@{}}
\toprule
System & Method & $J_1$  \\
\midrule
\multirow{2}{*}{\ch{La2CuO4}}
 & non-TC CCSD (DZ) &  101.5 \\
 & DMET+CC (DZ)~\cite{cui2022} & 106 \\

\bottomrule
\end{tabular}
\end{table}

To compare against the results of Cui \textit{et al.}, we computed first $J_1$ in the $4$-magnetic-site cluster with CCSD in DZ basis. Cui \textit{et al.} use def2-SVP basis, which is of similar quality. They also did not use ECPs like in this study. Table~\ref{tab:cui comparison} shows the evaluated $J_1$ with both workflows. It can be seen that our $J_1$ is $4.5$meV smaller. The difference is so small that it is hard to determine the cause in the discrepancy; slightly different basis, use of pseudopotentials, or different embedding. Nevertheless, the close similarity in the provided $J_1$ connects our workflow in this work to earlier studies and shows that use of ECPs, and ionic embedding, as opposed to all-electron calculations and DMET, have little effect.%

We did an analysis of finite-size errors in 2-magnetic-site clusters by comparing to results in 4-magnetic-site clusters. With 4-site clusters we can only use single-reference methods, as we estimate magnetic couplings according to Eq.~\ref{eq:inversion}. We compared the 4-site results to the 2-site results, both obtained with single-reference coupled cluster (i.e., with broken-symmetry reference determinant for the singlet), to get a balanced estimate.

\begin{table}[h]
\centering
\caption{Finite-size test of the embedding clusters: nearest-neighbour coupling $J_1$ (meV) from single-determinant DCSD, for the 2-site and the 4-site cluster of each material, with non-TC results presented in the cc-pVDZ (non-TC) and xTC-PP results presented in the composite TZ/DZ bases. $\Delta = J_1(\text{4-site}) - J_1(\text{2-site})$. The last column is the experimental value.}
\label{tab:4site}
\footnotesize
\setlength{\tabcolsep}{3.5pt}
\begin{tabular}{@{}llcccc@{}}
\toprule
System & Method & 2-site & 4-site & $\Delta$ & Exp. \\
\midrule
\multirow{2}{*}{\ch{La2CuO4}}
 & non-TC DCSD (DZ) & 109.0 & 111.6 & $+2.6$ & \multirow{2}{*}{$143 \pm 2$~\cite{Headings2010}} \\
 & xTC-PP DCSD (TZ/DZ) & 146 & 143 & $-3$& \\
\midrule
\ch{NdNiO2}
 & non-TC DCSD (DZ) &  69.2 &  65.5 & $-3.7$ & $63.6 \pm 3.3$~\cite{lu2021} \\
 & xTC-PP (TZ/DZ) &  87.2 &  81.1 & $-6.1$ & ($87.6^{*}$) \\
\bottomrule
\end{tabular}

\vspace{3pt}
{\footnotesize $^{*}$Hubbard-constrained $J_1$, derived from the fit of Lu \textit{et al.}~\cite{lu2021}; see Supplement~\cite{Supplement}.}
\end{table}

Table~\ref{tab:4site} shows the evaluated $J_1$ for $2$- and $4$-site clusters with single-determinant DCSD with non-TC and xTC-PP methods in DZ and TZ/DZ bases, respectively. The corrections by the larger clusters are from $-3$ to $-6$~meV in TZ/DZ basis with xTC-PP Hamiltonians. It should be also noted that, if we apply the xTC-PP correction for NdNiO$_2$ obtained in Table \ref{tab:4site} to the QZ/DZ $J_1$ from Table \ref{tab:2det_main} (where we show the 2det-DCSD result), we get a value of $86.1$~meV, with only a $1.5$~meV difference to the derived experimental $J_1$.

\section{Conclusions}

In this work we evaluated nearest-neighbour magnetic spin coupling parameters $J_1$ in a series of cuprate and nickelate materials using a combination of transcorrelated Hamiltonians and coupled cluster methods. We demonstrated that, in the correlated treatment, the coupling between orbital breathing and superexchange hopping is essential for an accurate description of the singlet state. Then we demonstrated that this coupling is expected to be captured by the $2$-determinant coupled cluster theory. Based on the accuracy against experiment, we believe this is indeed the case, since without proper description of breathing, the results would have been qualitatively wrong~\cite{Bogdanov2022}.

We applied the latest transcorrelated Hamiltonian theory to the problem of computing $J_1$ in strongly correlated materials. The transcorrelated Hamiltonian was constructed with a Jastrow factor optimized in the presence of the embedding point-charge cloud, and with the singular value decomposition-based virtual space compression. This approach led to a drastic reduction of the basis set requirements for accurate $J_1$ values. We found that double-$\zeta$ -level basis sets are insufficient, both in the non-TC and xTC-PP calculations, and that the use of very affordable composite triple-$\zeta$ basis with transcorrelated Hamiltonians is sufficient to converge $J_1$ to the CBS limit. Without transcorrelation, sometimes a full extrapolation to CBS is required, which is more expensive and less straightforward.

The $J_1$ across the four cuprate materials was captured with TZ/DZ 2det-xTC-PP-DCSD very accurately, with highly affordable calculations (see Appendix D for timings).  NdNiO$_2$ value from experiment~\cite{lu2021} was the effective $J_\text{eff}$, resulting from a free $2$-parameter fit, and lower than our computed value by $\approx 30\%$. However, by deriving the corresponding Hubbard-constrained $J_1$, we found an excellent agreement with our results. The work therefore demonstrates how wavefunction-based theoretical methods can efficiently be used to extract accurate magnetic parameters in strongly correlated materials, especially if combined with explicit correlation treatment such as transcorrelation. The methods presented in this study  can thus complement both experimental studies and theory based on lattice models.

The current study paves the way towards future work on magnetic interactions beyond the nearest-neighbour couplings. The methods developed here can be extended to larger clusters and more complex materials, providing a pathway for the accurate prediction of magnetic properties in a wide range of strongly correlated systems. Transcorrelation with more sophisticated embedding techniques, such as periodic mean field or density matrix embedding theory, can be used to further improve the accuracy of the calculations, and to explore the effects of different embedding environments on the magnetic interactions. 

Limitations of the current approach include the restriction to 2-determinant reference states, and for systems with more complex multireference nature yet more sophisticated correlation solvers are needed and are actively being developed. In any case, the robustness of transcorrelation in the alignment to non-TC CBS coupled cluster results is yet another milestone for TC studies, establishing the TC theory as a powerful tool in the research of strongly correlated systems with heavy elements. 

\clearpage

\begin{acknowledgments}
The authors used Claude Opus 5.1--5.5 (Anthropic) to help prepare the figures, to build the workflow for the final set of calculations based on initial implementation by the authors, to help derive and implement the perturbation-theory and coupled-cluster treatment of the model Hamiltonian, to assist the literature review of the experimental data, and to review the manuscript and the Supplemental Material. GitHub Copilot was used to develop the code for the CASINO Jastrow optimization with the point-charge cloud. The authors verified the figures, the theory and its implementation, the results, and the individual input and output files of the codes used, and take full responsibility for the content of this article.
\end{acknowledgments}

\appendix

\section{Theory}

\subsection{Evaluation of the transcorrelated Hamiltonian}
\label{sec:methods_tc}

Transcorrelation (TC)~\cite{boys1969} is based on a similarity transformation of the Hamiltonian with a real-space Jastrow correlator $\tau$ (whose form and optimization are described in Appendix B). Expanded via the Baker--Campbell--Hausdorff formula, the transformation can be expressed with a commutator expression as~\cite{cohen2019},
\begin{equation}
    \hat H^{\mathrm{TC}} = e^{-\tau} \hat H e^{\tau}
    = \hat H + \bigl[\hat H, \tau\bigr]
    + \tfrac{1}{2}\bigl[\bigl[\hat H, \tau\bigr], \tau\bigr] \,,
\label{eq:tc_simtransf}
\end{equation}
because only the kinetic-energy operator and, if present, the pseudopotential operator fail to commute with the multiplicative $\tau$. Separating the two contributions, the kinetic commutators define the two-body term $\hat K$ and the three-body term $\hat L$,
\begin{equation}
    \hat K + \hat L
    = \Bigl[\tfrac{1}{2}\textstyle\sum_p \nabla^2_p,\, \tau\Bigr]
    + \tfrac{1}{2}\Bigl[\bigl[\tfrac{1}{2}\textstyle\sum_p \nabla^2_p,\, \tau\bigr],\, \tau\Bigr] \,,
\label{eq:tc_KL}
\end{equation}
and the pseudopotential commutators the term
\begin{equation}
    \hat P
    = \Bigl[\textstyle\sum_p \hat V^{p}_{\mathrm{ECP}},\, \tau\Bigr]
    + \tfrac{1}{2}\Bigl[\bigl[\textstyle\sum_p \hat V^{p}_{\mathrm{ECP}},\, \tau\bigr],\, \tau\Bigr] \,,
\label{eq:tc_P}
\end{equation}
so that $\hat H^{\mathrm{TC}} = \hat H - (\hat K + \hat L) + \hat P$. For the all-electron Hamiltonian the expansion terminates identically at the second commutator; with (semi-)local pseudopotentials it is not guaranteed to terminate, but only the first two commutators need to be evaluated, and the three-body pseudopotential contributions are neglected under the assumption that three electrons rarely correlate simultaneously inside the ECP core region~\cite{simula2025ecp}.

The transformation preserves the eigenspectrum of $\hat H$ but renders $\hat H^{\mathrm{TC}}$ non-hermitian. In the xTC approximation~\cite{christlmaier2023}, the three-body term $\hat L$ is folded into effective zero-, one-, and two-body terms by contraction with the one-body reduced density matrix $\gamma$ of the reference, giving a second-quantized Hamiltonian of standard form,
\begin{equation}
    \hat H^{\mathrm{xTC}} = E^{\mathrm{xTC}}_0
    + \sum_{pq} h^{\mathrm{xTC}}_{pq}\, a^\dagger_p a_q
    + \tfrac{1}{2} \sum_{pqrs} V^{\mathrm{xTC}}_{pqrs}\, a^\dagger_p a^\dagger_q a_s a_r \,.
\label{eq:xtc_hamiltonian}
\end{equation}
The two-body xTC interaction is
\begin{align}
\label{eq:xtc_V}
    \begin{aligned}
    V^{\mathrm{xTC}}_{pqrs} &= V_{pqrs} + \Delta V_{pqrs} \,, \\
    \Delta V_{pqrs} &= -K_{pqrs} + P_{pqrs} + \Delta W_{pqrs} \,, \\
    \Delta W_{pqrs} &= -\sum_{tu}\bigl( L_{prtqsu} - L_{prtqus} - L_{prtusq} \bigr)\, \gamma_{tu} \,,
    \end{aligned}
\end{align}
where $V_{pqrs}$, $K_{pqrs}$, $P_{pqrs}$, and $L_{pqrstu}$ are the matrix elements of the Coulomb interaction and of the operators of Eqs.~\eqref{eq:tc_KL}--\eqref{eq:tc_P} in the orbital basis, and $\Delta W_{pqrs}$ is the xTC closure of the three-body term. The one-body and scalar terms acquire the corrections~\cite{christlmaier2023}
\begin{align}
\label{eq:xtc_h}
    \begin{aligned}
    h^{\mathrm{xTC}}_{pq} &= h_{pq} + \Delta h_{pq} \,, \\
    \Delta h_{pq} &= -\tfrac{1}{2}\sum_{tu}\bigl( \Delta W_{ptqu} - \Delta W_{ptuq} \bigr)\, \gamma_{tu} \,, \\
    E^{\mathrm{xTC}}_0 &= E_0 + \Delta E^{\mathrm{xTC}}_0 \,, \quad
    \Delta E^{\mathrm{xTC}}_0 = -\tfrac{1}{3}\sum_{tu} \Delta h_{tu}\, \gamma_{tu} \,.
    \end{aligned}
\end{align}
 Equations~\eqref{eq:xtc_hamiltonian}--\eqref{eq:xtc_h} constitute the complete recipe for the xTC(-PP) Hamiltonian; any second-quantized correlation solver that can treat non-hermitian Hamiltonians can process it.

The commutators in Eqs.~\eqref{eq:tc_KL} and~\eqref{eq:tc_P} involve derivatives of the Jastrow factor ($\nabla \tau$, $\nabla^2 \tau$) multiplying orbitals and orbital gradients, for which no closed-form integrals exist for a general $\tau$. The correction integrals $K_{pqrs}$, $P_{pqrs}$, and the $\gamma$-contractions of $L$ are therefore evaluated by numerical quadrature on atom-centred real-space grids: the orbitals, their gradients, and the Jastrow derivatives are tabulated on the grid points and contracted with the quadrature weights. Crucially, the six-index tensor $L_{pqrstu}$ is never stored; its contractions with $\gamma$ in Eq.~\eqref{eq:xtc_V} are carried out directly during the grid summation, which is the source of the favourable cost of the xTC scheme~\cite{christlmaier2023}. Throughout the main text, the prefix TC (or xTC-PP) on a method name indicates that the corresponding solver acts on the Hamiltonian of Eq.~\eqref{eq:xtc_hamiltonian}.

As we mention in the main text, we optimize the Jastrow factor for the singlet state. The similarity transformation preserves the eigenspectrum of the Hamiltonian, and thus the singlet Jastrow should also produce the triplet energies correctly. However, 1-RDM $\gamma$ used in the xTC approximation is still different for the singlet and triplet. The singlet has in the CAS(2,2) solution a charge-transfer channel that the triplet does not, leading to non-diagonal 1-RDM elements in the basis of localized magnetic orbitals. The non-diagonal elements are, however, very small, and we tested that using a state-specific $\gamma$ has only a negligible effect on the energetics. 

As recently shown~\cite{hauskrecht2026additive}, the correlation energy converges faster to CBS limit than the reference energy when using xTC Hamiltonians. Therefore we adopt the recently introduced virtual space compression scheme~\cite{cheng2026compression}, where the ROHF and Jastrow optimization are done in a large basis, here, the cc-pVQZ basis. After these steps, we use singular value decomposition to compress the size of the virtual space, while keeping the reference orbitals at cc-pVQZ quality, into the size of the chosen basis, such as the composite basis we use for non-TC calculations. See Appendix B for details on the choices on basis sets.

\section{Computational details}
\label{sec:comp_details}

\subsection{Ionic Embedding and the Atomic Fragments}

 The magnetic coupling $J_1$ is local, determined essentially by the two interacting magnetic centres and their immediate surroundings; hence it suffices to describe explicitly only a local fragment of the crystal, embedded in a field mimicking the rest of the lattice~\cite{deGraaf2016}. In the ionic embedding used in this work, the first shell of atoms surrounding the cluster is replaced by formal charges equipped with pseudopotentials, mimicking the Coulomb and exclusion effects of the immediate surroundings~\cite{martin1993}, while the rest of the lattice is represented by point charges whose values are assigned according to Evjen's method~\cite{Evjen1932} and which reproduce the Madelung potential of the crystal.

The fragments in Fig. \ref{fig:structures} are constructed from the periodic crystal structures, taken from experiments~\cite{hyatt2004,radaelli1994,sapina1990,hayward2003,shores2005}. By using the Chargedel-package~\cite{chargedel}, we construct the embedded fragments by picking the two active transition metal atoms and their nearest-neighbour ligands, and then adding the nearest-neighbour spacer cations: Sr, La, Nd, Li, and Zn. In addition, we pick the outer magnetic neighbour sites of the two central transition-metal atoms. The transition metal atoms that are not in the active sites, and the spacer atoms, are replaced by total-ion pseudopotentials without basis functions. See Fig.~\ref{fig:structures} for the atomic clusters used.

With the fragments established, we then construct the embedding point-charge cloud around the fragment. The point charges are placed on the atomic positions of the periodic crystal structure, and their values are set to the formal charges of the atoms, but after a predetermined number of nearest-neighbour periodic images of the fragment the charges are scaled down. The number of charges, and the number of shells of periodic images with explicit atomic charges, are parameters, grown until the produced charge cloud reproduces the Ewald-summation electrostatic potential of the periodic crystal structure to within $10^{-6}$~Ha of the Ewald-summation result. The final point-charge clouds contain 10\,612 charges for \ch{Sr2CuO3}, 35\,026 for \ch{La2CuO4}, 38\,162 for \ch{Li2CuO2}, 5\,920 for \ch{NdNiO2}, and 91\,706 for herbertsmithite.

\subsection{Basis sets and pseudopotentials}
\label{sec:basis_ecp}

All calculations remove chemically inert cores with pseudopotentials; the pseudopotential and basis-set assignments for every system are collected in \Cref{tab:basis_ecp}. Throughout, the magnetically active Cu and Ni atoms and the ligands O, Cl and H carry ccECP pseudopotentials~\cite{bennett2017,bennett2018} together with the matching correlation-consistent cc-pV$X$Z basis sets, $X=\mathrm{D},\mathrm{T},\mathrm{Q}$.  

For the periodic materials the composite bases used assign the larger $X$ to the two magnetic centres and the bridging oxygen and cc-pVDZ to the remaining ligands (in the case of xTC-PP Hamiltonians, these composite basis sets are assigned only to the virtual space, in the SVD-based virtual space compression scheme, see above). The surrounding spacer cations and non-active Cu or Ni atoms are described by bare large-core Stuttgart--Cologne (SC) pseudopotentials~\cite{fuentealba1985,fuentealba1982,igelmann1987}. In La$_2$CuO$_4$ clusters, we use for La$^{3+}$ the Ba pseudopotential with 54 frozen core electrons. For Nd$^{3+}$, we use the same Ba PP as for La; by setting the effective nuclear charge to $Z_{\mathrm{eff}}=3$, we remove the $f$ electrons from the valence space.  For non-active Cu atoms in the clusters (non-active), we use Zn PP with 28 electrons in frozen core~\cite{igelmann1987}. For non-active Ni atoms in the NdNiO$_2$ clusters, we use Cu-PP with 28 frozen core electrons~\cite{igelmann1987}. Sr PP from \cite{fuentealba1985} with 36 frozen core electrons is used in the Sr$_2$CuO$_3$ cluster for Sr$^{2+}$.

\begin{table*}[t]
\centering
\caption{Pseudopotentials and basis sets for each system. Active atoms and ligands carry ccECP pseudopotentials and the matching cc-pV$X$Z basis sets, $X=\mathrm{D},\mathrm{T},\mathrm{Q}$; non-active transition metals and spacer cations are represented by bare total-ion pseudopotentials without basis functions (---). $Z_{\mathrm{eff}}$ is the effective core charge of the pseudopotential. SC = Stuttgart--Cologne; where the SC pseudopotential is borrowed from a different element, that element is given in parentheses. The La$^{3+}$ and Nd$^{3+}$ spacers use the Ba ECP54SDF core relabelled to $Z_{\mathrm{eff}}=3$ (trivalent, no $f$ electrons in the valence space).}
\label{tab:basis_ecp}
\small
\begin{tabular}{@{}lllcl@{}}
\toprule
System & Atom & Pseudopotential & $Z_{\mathrm{eff}}$ & Basis set \\
\midrule
\multirow{4}{*}{\ch{Sr2CuO3}}
 & Cu (active)     & ccECP~\cite{bennett2017,bennett2018} & 19 & cc-pV$X$Z \\
 & O               & ccECP~\cite{bennett2017,bennett2018} & 6  & cc-pV$X$Z \\
 & Cu (non-active) & SC ECP28SDF (Zn)~\cite{igelmann1987} & 2 & --- \\
 & Sr (spacer)     & SC ECP36SDF~\cite{fuentealba1985} & 2 & --- \\
\midrule
\multirow{4}{*}{\ch{La2CuO4}}
 & Cu (active)     & ccECP~\cite{bennett2017,bennett2018} & 19 & cc-pV$X$Z \\
 & O               & ccECP~\cite{bennett2017,bennett2018} & 6  & cc-pV$X$Z \\
 & Cu (non-active) & SC ECP28SDF (Zn)~\cite{igelmann1987} & 2 & --- \\
 & La (spacer)     & SC ECP54SDF (Ba)~\cite{fuentealba1985} & 3 & --- \\
\midrule
\multirow{4}{*}{\ch{Li2CuO2}}
 & Cu (active)     & ccECP~\cite{bennett2017,bennett2018} & 19 & cc-pV$X$Z \\
 & O               & ccECP~\cite{bennett2017,bennett2018} & 6 & cc-pV$X$Z \\
 & Cu (non-active) & SC ECP28SDF (Zn)~\cite{igelmann1987} & 2 & --- \\
 & Li (spacer)     & ccECP~\cite{bennett2017,bennett2018} & 1 & --- \\
\midrule
\multirow{4}{*}{\ch{NdNiO2}}
 & Ni (active)     & ccECP~\cite{bennett2017,bennett2018} & 18 & cc-pV$X$Z \\
 & O               & ccECP~\cite{bennett2017,bennett2018} & 6 & cc-pV$X$Z \\
 & Ni (non-active) & SC ECP28SDF (Cu)~\cite{igelmann1987} & 1 & --- \\
 & Nd (spacer)     & SC ECP54SDF (Ba)~\cite{fuentealba1985} & 3 & --- \\
\midrule
\multirow{5}{*}{Herbertsmithite}
 & Cu (active)     & ccECP~\cite{bennett2017,bennett2018} & 19 & cc-pV$X$Z \\
 & O               & ccECP~\cite{bennett2017,bennett2018} & 6 & cc-pV$X$Z \\
 & Cl              & ccECP~\cite{bennett2017,bennett2018} & 7 & cc-pV$X$Z \\
 & H (cap)         & ccECP~\cite{bennett2017,bennett2018} & 1 & cc-pV$X$Z \\
 & Zn (spacer)     & SC ECP28SDF~\cite{igelmann1987} & 2 & --- \\
\bottomrule
\end{tabular}
\end{table*}

\subsubsection{Complete-basis-set estimate}

The non-TC CC values are extrapolated to the complete-basis-set (CBS) limit by
treating the reference and the correlation contribution separately, since the
two converge at different rates. Writing $E^{\sigma}_X$ for the total energy of the state
$\sigma\in\{S,T\}$ obtained with
the composite basis of cardinal number $X$ ($X=2,3,4$ for DZ, TZ/DZ and
QZ/DZ), the CAS(2,2) reference part is extrapolated with the three-point
exponential form
\begin{equation}
  E^{\mathrm{ref},\sigma}_{\mathrm{CBS}} = E^{\mathrm{ref},\sigma}_{4}
  - \frac{\left(E^{\mathrm{ref},\sigma}_{4}-E^{\mathrm{ref},\sigma}_{3}\right)^{2}}
         {\left(E^{\mathrm{ref},\sigma}_{4}-E^{\mathrm{ref},\sigma}_{3}\right)
          -\left(E^{\mathrm{ref},\sigma}_{3}-E^{\mathrm{ref},\sigma}_{2}\right)},
  \label{eq:cbs_ref}
\end{equation}
which is exact for a geometric decay of the increments. We retain the QZ/DZ value instead of $E^{\mathrm{ref},\sigma}_{\mathrm{CBS}}$ for Li$_2$CuO$_2$, because the reference energy change is only $1$--$2$~meV and the increments do not decay geometrically. The correlation part,
$E^{\mathrm{corr},\sigma}_{X}=E^{\sigma}_{X}-E^{\mathrm{ref},\sigma}_{X}$,
decays as $X^{-3}$ and is
extrapolated from the two largest bases,
\begin{equation}
  E^{\mathrm{corr},\sigma}_{\mathrm{CBS}}
  = \frac{4^{3}E^{\mathrm{corr},\sigma}_{4}-3^{3}E^{\mathrm{corr},\sigma}_{3}}{4^{3}-3^{3}}
  = \frac{64\,E^{\mathrm{corr},\sigma}_{4}-27\,E^{\mathrm{corr},\sigma}_{3}}{37}.
  \label{eq:cbs_corr}
\end{equation}
Because the composite bases carry a DZ description on the ligand atoms other than the bridging oxygen, a
composite-basis correction is added on top, taken as the difference between
the full triple-zeta and the composite TZ/DZ result at the same level of
theory,
\begin{equation}
  \Delta^{\sigma}_{\mathrm{TZ}}
  = E^{\sigma}(\mathrm{TZ}) - E^{\sigma}(\mathrm{TZ/DZ}),
  \label{eq:cbs_delta}
\end{equation}
so that the CBS total energy of each state is
$E^{\sigma}_{\mathrm{CBS}} = E^{\mathrm{ref},\sigma}_{\mathrm{CBS}}
+ E^{\mathrm{corr},\sigma}_{\mathrm{CBS}} + \Delta^{\sigma}_{\mathrm{TZ}}$,
and the coupling follows as
$J_{1,\mathrm{CBS}} = E^{T}_{\mathrm{CBS}} - E^{S}_{\mathrm{CBS}}$.
The transcorrelated
results are not extrapolated: they are reported at the basis at which they were
computed.

\subsection{Orbital generation}
\label{sec:orbgen_jastrow}

Reference orbitals are generated in three steps: (i) a restricted open-shell Hartree--Fock calculation for the $m_s=1$ triplet state; (ii) a CASSCF calculation in the chosen active space of two active Cu 3d$_{x^2-y^2}$  orbitals; and (iii) separate Pipek--Mezey~\cite{pipek1989} localizations of the occupied, active, and virtual orbital spaces. We chose the localized, CASSCF-optimized Cu 3$d_{x^2-y^2}$ orbitals of both active sites, obtained with given active space size, and construct the singlet reference as a linear combination of the two open-shell determinants as explained in the main text.

\subsection{Jastrow optimization}

The Jastrow correlator entering Eq.~\eqref{eq:tc_simtransf} is of the Drummond--Towler--Needs form~\cite{drummond2004}, containing two-body ($u$), one-body ($\chi$), and electron--electron--nucleus three-body ($f$) terms, each a polynomial with a finite cutoff radius; the $u$ term enforces the electron--electron cusp condition. Since all calculations employ pseudopotentials, no electron--nucleus cusp treatment is required. The variational parameters are optimized in variational Monte Carlo with the CASINO package~\cite{needs2020} by minimizing the variance of the transcorrelated reference energy~\cite{haupt2023}. For this work we modified the CASINO source code to evaluate the electronic potential energy in the electrostatic field of the point-charge cloud at every Metropolis-sampled electronic configuration, so that the Jastrow factor is optimized directly for the embedded-cluster model rather than for the bare cluster. This simple implementation is inefficient: to avoid serial correlation of the samples, the code runs a decorrelation loop of 50 Metropolis steps between successive energy evaluations, yet the point-charge potential is currently evaluated at every Metropolis step. Evaluating it only at the energy-evaluation steps would accelerate the Metropolis sampling by a factor of 50; the parameter-optimization step is unaffected, but the overall cost of the VMC stage could still be reduced by roughly an order of magnitude.

\section{Workflow}

\begin{figure*}[t]
     \centering
        \includegraphics[width=0.9\textwidth]{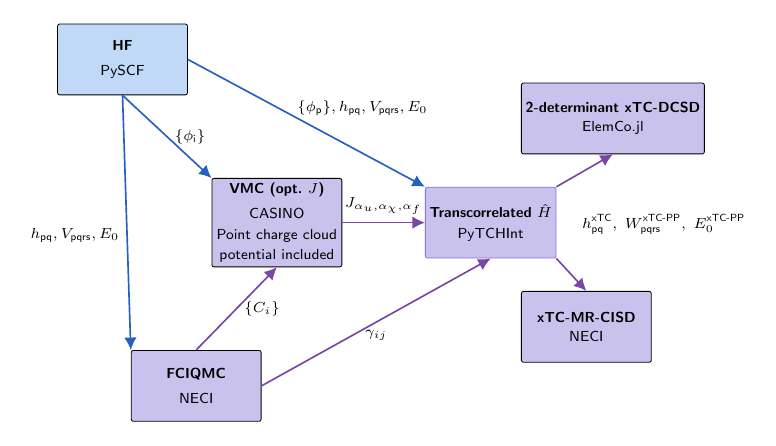}
        \caption{Computational workflow for spin coupling calculations, illustrating the sequence of methods and data flow between them. $C_i$ are determinant coefficients from FCIQMC, $\phi$ are HF molecular orbitals, $h_{pq}$ and $V_{pqrs}$ are one- and two-electron integrals, $E_0$ is the reference energy, and $\gamma_{ij}$ is the one-body reduced density matrix. Figure prepared with AI assistance and verified by the authors; see Acknowledgments.}
        \label{fig:workflow}
\end{figure*}

The complete computational workflow, from the reference orbitals to the correlated singlet--triplet gaps, is summarized in Fig.~\ref{fig:workflow}. The reference orbitals -- ROHF, CASSCF, and the subsequent localization -- are obtained with PySCF~\cite{sun2020}. NECI~\cite{guther2020} is then used to perform a CASCI calculation in the CAS(2o,2e) active space, which yields the 2-determinant reduced density matrix required for xTC Hamiltonians, computed by PyTCHInt, and the determinantal coefficients of the VMC trial wave function. This step could be carried out solely in PySCF, but the present workflow is scalable to arbitrary active-space sizes: for large active spaces the CASCI step can be performed efficiently with the FCIQMC machinery of NECI. The Jastrow parameters are optimized in variational Monte Carlo with the CASINO package~\cite{needs2020}. Given the optimized Jastrow factor, the reference orbitals, and the reduced density matrix, the xTC and xTC-PP correction integrals of Eqs.~\eqref{eq:xtc_V} and~\eqref{eq:xtc_h} are evaluated numerically with our in-house code PyTCHInt. The resulting non-TC and xTC second-quantized Hamiltonians are passed via the FCIDUMP format~\cite{Knowles1989} or as NumPy binary files~\cite{Harris2020} to the ElemCo.jl package~\cite{kats2026elemco}, which performs the 2det-CCSD and 2det-DCSD calculations.

Separately from the workflow, the non-TC 2det-CC methods are done directly with the orbitals from the PySCF phase. 

\section{Timings}
\label{sec:timings}

The computational cost of the transcorrelated workflow is dominated by the VMC Jastrow optimization, which is run on eight 128-core nodes; the reference generation, the xTC integral evaluation, and the 2det-DCSD solution are comparatively cheap. \Cref{tab:timings} breaks down, for each embedded-cluster system, the cost of the TC chain behind the xTC-PP TZ/DZ results -- cc-pVQZ reference and Jastrow optimization, followed by xTC integral evaluation and 2det-DCSD in the compressed TZ/DZ virtual space -- in AMD 128-core node-hours. The complete chain for all five systems amounts to 392 node-hours. For scale, the last column lists the cost of the full non-TC survey of Fig.~\ref{fig:2det_comparison} (reference plus the six CC variants in each basis from DZ to QZ/DZ), 1235 node-hours in total. The reported node-hours are approximate: the runs span several node generations, and part of the VMC timings are reconstructed from CPU times. The dominance of the VMC step is likely due to the present, inefficient implementation of the point-charge potential evaluation during the Metropolis sampling; see the Orbital generation and Jastrow optimization sections in Appendix B.

\begin{table}[t]
\centering
\caption{Computational cost of the TC workflow for each embedded cluster, in AMD 128-core node-hours (walltime $\times$ nodes): reference generation at cc-pVQZ (Ref.), VMC Jastrow optimization on eight nodes (Jastrow), xTC integral evaluation (xTC), and 2det-DCSD in the compressed TZ/DZ space. The last column gives the total cost of the non-TC basis-set survey (reference plus six CC variants per basis, DZ to QZ/DZ). For herbertsmithite only the first of the two Jastrow-optimization stages was recorded, and for Li$_2$CuO$_2$ neither (`--'); the affected totals are lower bounds. Values are approximate, as the runs span several node generations; totals are computed from unrounded timings.}
\label{tab:timings}
\footnotesize
\setlength{\tabcolsep}{4pt}
\begin{tabular}{@{}lrrrrrr@{}}
\toprule
System & Ref. & Jastrow & xTC & 2det-DCSD & TC total & non-TC \\
\midrule
Sr$_2$CuO$_3$   & 3.5 &  22.2 & 2.4 & 4.7 &  32.8 &  113.9 \\
La$_2$CuO$_4$   & 2.9 & 117.4 & 8.2 & 5.7 & 134.1 &  419.2 \\
Li$_2$CuO$_2$   & 1.5 &    -- & 1.8 & 4.4 &   7.7 &   93.1 \\
NdNiO$_2$       & 1.9 &  41.3 & 4.4 & 4.2 &  51.9 &  104.1 \\
Herbertsmithite & 4.0 & 146.4 & 5.1 & 9.6 & 165.1 &  504.3 \\
\midrule
Total           & 13.8 & 327.3 & 21.9 & 28.6 & 391.5 & 1234.7 \\
\bottomrule
\end{tabular}
\end{table}

\bibliography{references}

\end{document}